\documentclass[a4paper,scale=0.8,onecolumn,nobibnotes,nofootinbib]{revtex4}

\usepackage{amsmath,amssymb}
\usepackage{graphicx,subfigure,epsfig}
\usepackage{hyperref}
\usepackage{geometry}
\hypersetup{
    colorlinks=true,
    citecolor=blue,
    linkcolor=red,
    filecolor=magenta,
    urlcolor=cyan,}

\newcommand{\be}{\begin{equation}}
\newcommand{\ee}{\end{equation}}

\begin{document}

\title{The study of the generalized  Klein-Gordon oscillator in the context of the Som-Raychaudhuri space-time}
\author{Lan Zhong$^{1}$,Hao Chen$^{1}$, Hassan Hassanabadi\footnote{ h.hasanabadi@shahroodut.ac.ir}$^{2}$, Zheng-Wen Long\footnote{ zwlong@gzu.edu.cn}$^{1}$ and Chao-Yun Long$^{1}$}
\affiliation{$1$ College of Physics, Guizhou University, Guiyang, 550025, China.\\
 ${2}$ Department of Physics, Shahrood University of Technology, P.O. Box 3619995161-316, Shahrood, Iran.
}

\date{\today}

\begin{abstract}
  \textbf{Abstract}. In this paper we study the relativistic scalar particle described by the Klein-Gordon interacts with the uniform magnetic field in the context of the Som-Raychaudhuri space-time. Based on the property of the biconfluent Heun function equation, the corresponding Klein-Gordon oscillator and generalized Klein-Gordon oscillator under considering the Coulomb potential are separately investigated, and the analogue of the Aharonov-Bohm effect is analyzed in this scenario. On this basis, we also give the influence of different parameters including the parameter $\alpha$ and oscillator frequency $\omega$, and the potential parameter $\xi_{2}$ on the energy eigenvalues of the considered systems.
\end{abstract}
\maketitle
~~~~~~~~~\textsl{Keywords:} Klein-Gordon oscillator, Som-Raychaudhuri space-time, Biconfluent Heun function equation, Coulomb potential.

\section{Introduction}
The vacuum phase transition in the early universe may have produced different types of topological defects such as domain walls\cite{1}, global monopole\cite{2} and cosmic string. Cosmic string change the topology of the medium and have been receiving extensive attention\cite{47,48,49} as one of the space-time where topological defects arise. G$\ddot{o}$del\cite{3} obtained one solution from Einstein equations and it has been well-known as the first cosmological solution for the rotating matter and got further development in Refs\cite{4,5}. There are a few numbers of  special cases about G$\ddot{o}$del-type space-time, when the curvature is zero, it reduced to the Som-Raychaudhuri(SR) space-time. Recent years, the investigations relevant about the G$\ddot{o}$del-type space-time and Som-Raychaudhuri space-time with relativistic quantum systems have been one hot subject. Hassan Hassanabadi et al\cite{6} considered DKP equation and DKP oscillator in SR space-time, got wave function by the generalized Laguerre polynomial and analyzed some effects to energy density from parameters and so on.\\
Dirac oscillator\cite{10}, which is a few model can be solved exactly, has always attracted many interests. Moshinsky and Szczepaniak firstly taken substitution term $\vec{p}$+im$\omega r\hat{r}$ into momentum operator in the Dirac equation and called this equation Dirac oscillator. The studies about Dirac oscillator have involved many fields such as Foldy-Wouthuysen transformation\cite{11}, symmetry Lie algebra\cite{12} and more about the Dirac equation and Dirac oscillator in Refs\cite{49,13,14,15,16,17,18,19,20,22}. The Dirac equation and Dirac oscillator describe the motion of spin-$\frac{1}{2}$ particles, while the Klein-Gordon equation plays an important role in solving the problem about spin-0 particles. In fact, inspired by Dirac oscillator, taking the same measures, an analogous Dirac oscillator coupling is introduced into the Klein-Gordon equation by Bruce and Minning\cite{23}, the Klein-Gordon equation becomes the Klein-Gordon oscillator. The K-G equation and K-G oscillator have always attracted attention by scholars and studied in many fields, including: noncommutative space\cite{24}, noncommutative phase space\cite{25}, EUP \cite{26}, Kaluza Klein theory \cite{27,28}and more in\cite{29,30,31,32,33}. The K-G equation and K-G oscillator with different physical potentials have be solved, including: Coulomb potential\cite{34}, Kratzer potential\cite{35}, generalized Hulthen potential\cite{36}. If the momentum term $\vec{p}$+im$\omega r\hat{r}$ is replaced by $\vec{p}$+im$\omega f(r)\hat{r}$, where $f(r)$ is the function about r, the systems in which this term exists are often referred to as generalized. In the cosmic string space-time, generalized K-G oscillator\cite{37} and generalized Kemmer oscillator\cite{38} were successively studied and the investigation about the generalized DKP oscillator under the chiral conical space-time finished in Ref\cite{39}. \\
Our work is motivated by the spin-0 massive charged particle with Coulomb potential in the context of the topologically trivial flat G\"odel-type space-time\cite{tt1} and the Klein-Gordon oscillator in the Som-Raychaudhuri space-time\cite{g1}, and analyze the analogue of the Aharonov-Bohm effect \cite{t2}, the similarly studies in the Refs\cite{t3,t1,8,9}. Therefore, it seems interesting to study of the generalized  Klein-Gordon oscillator in the context of the Som-Raychaudhuri space-time. The structure of this essay is as follows. In the next section, we briefly review the generalized K-G oscillator in the Som-Raychaudhuri space-time. In Section 3 we derive the energy spectrum of  the K-G oscillator and generalized K-G oscillator with Coulomb potential under an external magnetic field in Som-Raychaudhuri space-time. Eventually, the conclusion is presented in Section 4.
\section{The general expression of the generalized  Klein-Gordon oscillator in the Son-Raychaudhuri space-time}
The G$\ddot{o}$del-type solution with torsion and a topological defect can be written as the  space-time metric  in cylindrical coordinates is given by\cite{3}
\begin{equation}
d s^{2}=-(d t+\alpha\Omega \frac{\sin h^{2}lr}{l^{2}} d \phi)^{2}+\alpha^{2}\frac{\sin h^{2}2lr}{4l^{2}}d \phi^{2}+d r^{2}+d z^{2},
\end{equation}
with $0\leq \phi\leq 2\pi$, $0\leq r$, $-\infty\leq z \leq\infty$, respectively. While the parameter $\Omega$ characterizes the vorticity of the space-time. (We set $c=1$, $\hbar=1$). The angular parameter $\alpha$ with range values 0 to 1 has relation with $\eta$, shows $\alpha=1-4\eta$, of which $\eta$ is the linear mass density.
For the metric Eq.(1), when the $l\rightarrow 0$, it reduces to the well-knew Som-Raychaudhuri space-time and rewrites as \cite{3,40}
\begin{equation}
d s^{2}=-(d t+\alpha\Omega r^{2} d \phi)^{2}+\alpha^{2}r^{2}d \phi^{2}+d r^{2}+d z^{2}.
\end{equation}
Note this metric is of Lorentzian with $(-,+,+,+)$ or $+2$ signature. From the line element $Eq. (2)$, the related quantities can be obtained one by one, such as the $g=det(g_{\mu\nu})=-\alpha^{2}r^{2}$, the contravariant components $g^{\mu \nu}$ of contravariant components $g_{\mu\nu}$
\begin{equation}g^{\mu \nu}=\left(\begin{array}{cccc}
r^{2}\Omega^{2}-1 & 0 & \frac{-\Omega}{\alpha} & 0 \\
0 & 1 & 0 & 0 \\
\frac{-\Omega}{\alpha} & 0 & \frac{1}{\alpha^{2}r^{2}} & 0 \\
0 & 0 & 0 & 1
\end{array}\right).\end{equation}
The K-G equation describes the motion of the charged scalar particles with mass M is \cite{41}
\begin{equation}\label{eq1}
[\frac{1}{\sqrt{-g}}D_{\mu}(\sqrt{-g}g^{\mu\nu}D_{\nu})]\Psi=M^{2}\Psi,
\end{equation}
with $D_{\mu}$ = $\partial_{\mu}-ieA_{\mu}$. Here, e represents electric charge, $A_{\mu}$ is four-vector potential about electromagnetic with expression $A_{\mu}=(0,0,A_{\phi},0)$. The work in the Ref \cite{51}, the authors investigate the spin-half particle with the combination magnetic field and Aharonov-Bohm potential which showed $B=B_{z,1}+B_{z,2}$, where $B_{z,1}=B$ and $B_{z,2}=\frac{\phi}{e}\frac{\delta(r)}{\gamma r}$ represent the uniform magnetic field and Aharonov-Bohm effect, respectively. And the corresponding vector potentials (in the Coulomb gauge) can be got. Motivated by \cite{51,43}, in this essay, $A_{\phi}$ is chosen to be
\begin{equation}\label{eq2}
A_{\phi}=-\frac{1}{2}\alpha B_{0}r^{2}+\frac{\Phi_{B}}{2\pi},
\end{equation}
where $\Phi_{B}=const$ is the internal magnetic quantum flux.
Introducing $M(r)=M+S(r)$ as the mass term, of which $S(r)$ is one scalar potential, one obtains
\begin{equation}\label{eq1}
[\frac{1}{\sqrt{-g}}D_{\mu}(\sqrt{-g}g^{\mu\nu}D_{\nu})]\Psi=(M+S(r))^{2}\Psi,
\end{equation}
If the momentum operator does transformation: $\vec{p_{\mu}}\rightarrow \vec{p_{\mu}}+iM\omega X_{\mu}$, where $X_{\mu}=(0,f(r),0,0)$, $f(r)$ is unknown function about r, $\omega$ is the frequency of oscillator, the K-G equation becomes the generalized K-G oscillator reads
\begin{equation}
\frac{1}{\sqrt{-g}}(D_{\mu}+M\omega X_{\mu})(\sqrt{-g}g^{\mu\nu}(D_{\nu}-M\omega X_{\nu}))\Psi=(M+S(r))^{2}\Psi.
\end{equation}
Combination the above equation, $Eq.(3)$ and $g=det(g_{\mu\nu})=-\alpha^{2}r^{2}$, there is
\begin{equation}\begin{array}{l}
[-\frac{\partial^{2}}{\partial t^{2}}+\{r\Omega \frac{\partial}{\partial t}-\frac{1}{\alpha r}(\frac{\partial}{\partial\phi } -ieA_{\phi})\}^{2}+\frac{\partial^{2}}{\partial r^{2}}+\frac{1}{r}\frac{\partial}{\partial r}-\frac{M\omega}{r}f(r)-M\omega f^{'}(r)-M^{2}\omega^{2}f^{2}(r)\\+\frac{\partial^{2}}{\partial z^{2}}-(M+S(r))^{2}]\Psi=0.
\end{array}\end{equation}
\section{The generalized K-G oscillator $f(r)$ is considered two different potential function }
\subsection{the Linear potential}
If set $f(r)=r$, the generalized K-G oscillator reduces to the K-G oscillator. By considering the Coulomb potential $\frac{\lambda}{r}$ as $S(r)$, we get
\begin{equation}\begin{array}{l}
[-\frac{\partial^{2}}{\partial t^{2}}+\{r\Omega \frac{\partial}{\partial t}-\frac{1}{\alpha r}(\frac{\partial}{\partial\phi } -ieA_{\phi})\}^{2}+\frac{\partial^{2}}{\partial r^{2}}+\frac{1}{r}\frac{\partial}{\partial r}-2M\omega
-M^{2}\omega^{2}r^{2}+\frac{\partial^{2}}{\partial z^{2}}-(M+\frac{\lambda}{r})^{2}]\Psi=0.
\end{array}\end{equation}
It is obvious that the metric is independent of $t,\varphi \ and \ z$ so we choose the below expression as possible solution
\begin{equation}
\Psi=e^{-iEt+il\phi+ikz}s(r),
\end{equation}
where $l=0,\pm1,\pm2,...,$ and $-\infty < k < +\infty$. Putting $Eq.(10)$ into $Eq.(9)$, the second order differential equation about $r$ reads
\begin{equation}
s^{''}+\frac{1}{r}s^{'}+[a_{4}-a_{1}\frac{1}{r^{2}}-a_{2}^{2}r^{2}-2a_{3}\frac{1}{r}]s=0,
\end{equation}
with new parameters
\begin{equation}\begin{array}{l}
a_{1}=\frac{l^{2}}{\alpha^{2}}+\frac{e^{2}\Phi_{B}^{2}}{4\alpha^{2}\pi^{2}}+\lambda^{2}-\frac{el\Phi_{B}}{\alpha^{2} \pi},\\
\\
a_{2}^{2}=E^{2}\Omega^{2}+e\Omega B_{0}E+\frac{e^{2}B_{0}^{2}}{4}+M^{2}\omega^{2},\\
\\
a_{3}=M\lambda,\\
\\
a_{4}=E^{2}+\frac{\Omega e \Phi_{B}}{\alpha \pi}E-\frac{2\Omega l}{\alpha}E-\frac{e B_{0} l}{\alpha}+\frac{e^{2}\Phi_{B} B_{0}}{2\pi \alpha}-2M \omega-k^{2}-M^{2}.
\end{array}\end{equation}
Take the below ansatz
\begin{equation}
s(r)=\exp(-\frac{1}{2}a_{2}r^{2})r^{\sqrt{a_{1}}}H(r),
\end{equation}
where $H(r)$ is to be determined, substituting $Eq.(13)$ into $Eq.(11)$, we have
\begin{equation}
H^{''}+((2\sqrt{a_{1}}+1)\frac{1}{r}-2a_{2}r)H^{'}+(a_{4}-2a_{2}-2\sqrt{a_{1}}a_{2}-2a_{3}\frac{1}{r})H=0.
\end{equation}
We take the change of variable $\rho=\sqrt{a_{2}}r$. Hence, $Eq.(14)$ is rewritten as
\begin{equation}\begin{array}{l}
H^{''}+(\frac{1+2\sqrt{a_{1}}}{\rho}-2\rho)H^{'} +[\frac{a_{4}}{a_{2}}-2-2\sqrt{a_{1}}-\frac{2a_{3}}{\sqrt{a_{2}}}\frac{1}{\rho}]H=0.
\end{array}\end{equation}
This second order differential equation is called as the biconfluent Heun equation\cite{44,53} and the related function named the biconfluent Heun function\cite{45,46}. Using the Frobenius method, the $H(\rho)$ can be shown by the power series expansion
\begin{equation}
H(\rho)=\Sigma_{n=0}C_{n}\rho^{n+j},
\end{equation}
by taking the power series expansion into the $Eq.(15)$, we get
\begin{equation}\begin{array}{l}
j(j+2\sqrt{a_{1}})C_{0}\rho^{j-2}+[(j+1)(j+1+2\sqrt{a_{1}})C_{1}-\frac{2a_{3}}{\sqrt{a_{2}}}C_{0}]\rho^{j-1}+\Sigma_{n=0}(n+j+2)(n+j+2+2\sqrt{a_{1}})C_{n+2}\rho^{n+j}
\\+\Sigma_{n=0}(\frac{a_{4}}{a_{2}}-2-2\sqrt{a_{1}}-2(n+j))C_{n}\rho^{n+j}-\Sigma_{n=0}\frac{2a_{3}}{\sqrt{a_{2}}}C_{n+1}\rho^{n+j}=0.
\end{array}\end{equation}
From the coefficients of $\rho^{j-2}, \rho^{j-1}$ and $\rho^{n+j}$, there are
\begin{equation}\begin{array}{l}
 j=0, \\or\\ j=-2\sqrt{a_{1}},
\end{array}\end{equation}
\begin{equation}
 C_{1}=\frac{2a_{3}/\sqrt{a_{2}}}{(2\sqrt{a_{1}}+1)(j+1)}C_{0},
\end{equation}
\begin{equation}
C_{n+2}=\frac{2a_{3}/\sqrt{a_{2}}}{(n+j+2)(n+j+2+2\sqrt{a_{1}})}C_{n+1}+\frac{2+2(n+j)+2\sqrt{a_{1}}-\frac{a_{4}}{a_{2}}}{(n+j+2)(n+j+2+2\sqrt{a_{1}})}C_{n}.
\end{equation}
In this essay, we only consider the case of $j=0$. So $Eq.(20)$ becomes
\begin{equation}
C_{n+2}=\frac{2a_{3}/\sqrt{a_{2}}}{(n+2)(n+2+2\sqrt{a_{1}})}C_{n+1}+\frac{2+2n+2\sqrt{a_{1}}-\frac{a_{4}}{a_{2}}}{(n+2)(n+2+2\sqrt{a_{1}})}C_{n},
\end{equation}
with $C_{0}=1$, we find the power series expansion turns into the polynomial expression of degree n when the following two conditions are true
\begin{equation}
C_{n+1}=0, \\\  \frac{a_{4}}{a_{2}}-2-2\sqrt{a_{1}}=2n,
\end{equation}
where n is a positive integer number. Combination the second condition and $Eq.(12)$, one obtains
\begin{equation}\begin{array}{l}
(2n+2+2\sqrt{\frac{1}{\alpha^{2}}\tilde{l}_{ef}^{2}+\lambda^{2}}) \sqrt{E^{2}\Omega^{2}+e\Omega B_{0}E+\frac{e^{2}B_{0}^{2}}{4}+M^{2}\omega^{2}}=E^{2} -\tilde{l}_{ef}(\frac{2\Omega E}{\alpha}+\frac{eB_{0}}{\alpha})-2M \omega-k^{2}-M^{2}.
\end{array}\end{equation}
 The expression of the energy levels about the K-G oscillator with Coulomb potential under an uniform magnetic field in the SR space-time is derived. We can know that the effective angular momentum $\tilde{l}_{ef}=l-\frac{e\Phi_{B}}{2\pi}$ modifies the energy spectrum, which cause the analogue effect to the Aharonov-Bohm effect\cite{t2,t3,D12,D13} with the relationship  $E_{n, \frac{\tilde{l}_{ef}}{\alpha}}\left(\Phi_{B}+\Phi_{0}\right)=E_{n, \frac{\tilde{l}_{ef}}{\alpha} \mp \tau}(\Phi_{B})$, $\Phi_{0}=\pm\frac{2\pi}{e}\tau$ with $\tau$ take values(1,2,3,4). Comparing $Eq.(20)$ and the $Eq.(29)$ which is from the Ref\cite{t1}, they have similar expression composition and they are equal when $\omega\rightarrow 0, \lambda\rightarrow 0, B_{0}\rightarrow 0$ of $Eq.(23)$ and $\xi\rightarrow 0,\xi_{c}\rightarrow 0 $ of $Eq.(29)$ from the Ref\cite{t1}. To analyze the above result, we respectively plot the positive energy eigenvalues versus the angular parameter $\alpha$ and oscillator frequency $\omega$, the vorticity parameter $\Omega$ and the Coulomb potential parameter $\lambda$. First, in the $Fig. 1$ , there are plots about the energy eigenvalues via the angular parameter $\alpha$ and oscillator frequency $\omega$. On the one hand, (a) directly shows that the energy decreases with the increase of $\alpha$ in cases of different values of the quantum
number, on the other hand, there is increase trend of E with the add values of $\omega$ in (b). In the $Fig. 2$, both (a) and (b) show the increasing trend of energy with $\Omega$ and $\lambda$, respectively.
\begin{figure}
\begin{minipage}[t]{0.45\linewidth}
\centerline{\includegraphics[width=7.5cm]{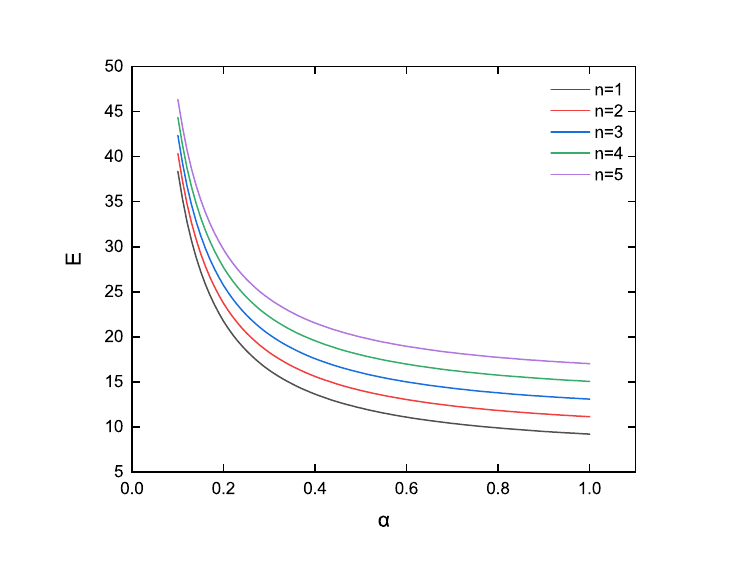}}
\centerline{(a)}
\end{minipage}
\begin{minipage}[t]{0.48\linewidth}
\centerline{\includegraphics[width=7.5cm]{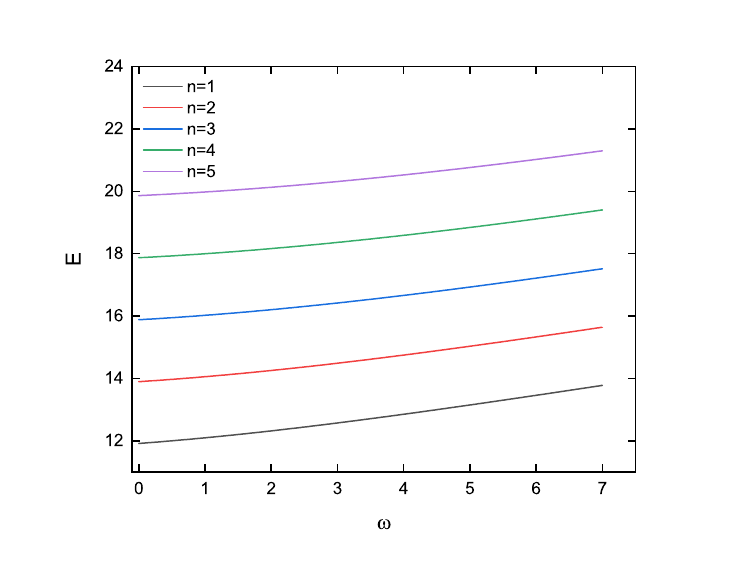}}
\centerline{(b)}
\end{minipage}
\parbox[c]{15.0cm}{\footnotesize{\bf Fig.~1.} (a) Energy eigenfunctions $E$ for different  angular parameter $\alpha$ with values of the quantum \\ number n(1, 2, 3, 4, 5), $M=e=\Phi_{B}=\omega=\lambda=B_{0}=l=\Omega=k=1$; \\
(b) Energy eigenfunctions $E$ fordifferent oscillator frequency $\omega$ with values of the quantum number \\n(1, 2, 3, 4, 5), $M=e= \Phi_{B}=\lambda=B_{0} =l=\Omega=k=1$, $\alpha=0.5$.}
\end{figure}
\begin{figure}
\begin{minipage}[t]{0.45\linewidth}
\centerline{\includegraphics[width=7.5cm]{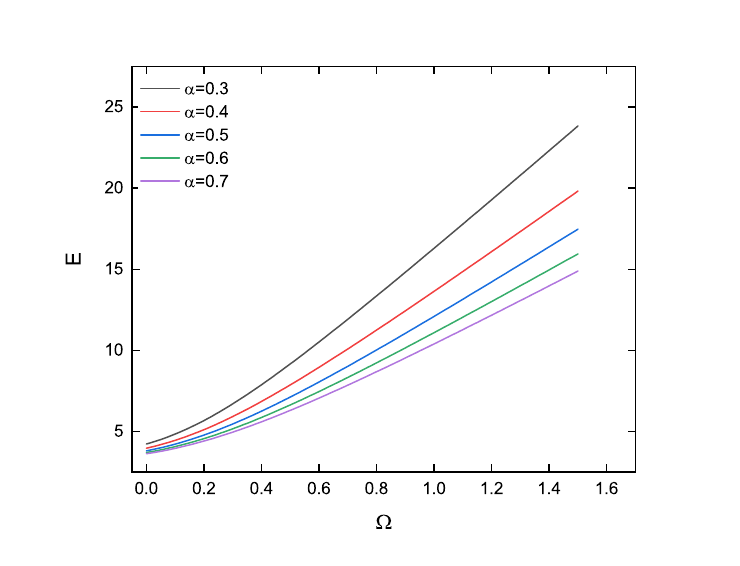}}
\centerline{(a)}
\end{minipage}
\begin{minipage}[t]{0.48\linewidth}
\centerline{\includegraphics[width=7.5cm]{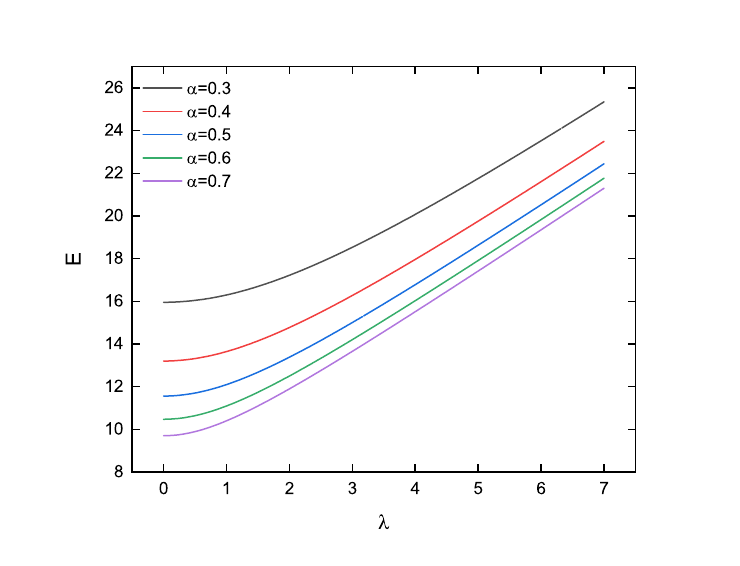}}
\centerline{(b)}
\end{minipage}
\parbox[c]{15.0cm}{\footnotesize{\bf Fig.~2.} (a) Energy eigenfunctions $E$ for different vorticity parameter $\Omega$ with values of the angular \\ parameter $\alpha$(0.3, 0.4, 0.5, 0.6, 0.7), $M=e=\Phi_{B}=n=\lambda=B_{0}=l=\omega=k=1$; \\
(b) Energy eigenfunctions $E$ for Coulomb-type potential parameter $\lambda$ with values of the angular \\parameter $\alpha$(0.3, 0.4, 0.5, 0.6, 0.7), $M=e=\Phi_{B}=n=B_{0}=\omega=l=\Omega=k=1$.}
\end{figure}

\subsection{the Cornell potential}
The second potential choice is the Cornell potential. Cornell potential\cite{D13,50} $\xi_{1}r+\frac{\xi_{2}}{r}$ appears in the study of quark systems, which can be seen as the combination of linear potential  and the Coulomb potential. We set $f(r)=\xi_{1}r+\frac{\xi_{2}}{r}$ while $S(r)$ is still Coulomb potential. With those, $Eq.(8)$ has new form as follows
\begin{equation}\begin{array}{l}
[-\frac{\partial^{2}}{\partial t^{2}}+\{r\Omega \frac{\partial}{\partial t}-\frac{1}{\alpha r}(\frac{\partial}{\partial\phi } -ieA_{\Phi})\}^{2}+\frac{\partial^{2}}{\partial r^{2}}+\frac{1}{r}\frac{\partial}{\partial r}-2M\omega \xi_{1}-M^{2}\omega^{2}\xi_{1}^{2}r^{2}\\-2M^{2}\omega^{2}\xi_{1}\xi_{2}-\frac{M^{2}\omega^{2}\xi_{2}^{2}}{r^{2}}+\frac{\partial^{2}}{\partial z^{2}}-(M+\frac{\lambda}{r})^{2}]\Psi(t,r,\varphi, z)=0,
\end{array}\end{equation}
where $\Psi(t,r,\varphi, z)$ are independent. Let us set
\begin{equation}
  \Psi=e^{-iEt+il\phi+ikz}s(r),
\end{equation}
with $l=0,\pm1,\pm2,...,$ and $-\infty < k < +\infty$. Inserting the $Eq.(25)$ into $Eq.(24)$, one obtains
\begin{equation}
s^{''}+\frac{1}{r}s^{'}+(b_{4}-b_{1}\frac{1}{r^{2}}-b_{2}^{2}r^{2}-2b_{3}\frac{1}{r})s=0,
\end{equation}
where $b_{1}, b_{2}, b_{3}, b_{4}$ as new parameters are given by
\begin{equation}\begin{array}{l}
b_{1}=M^{2}\omega^{2}\xi_{2}^{2}+\frac{l^{2}}{\alpha^{2}}+\frac{e^{2}\Phi_{B}^{2}}{4\alpha^{2}\pi^{2}}+\lambda^{2}-\frac{el\Phi_{B}}{\alpha^{2} \pi},\\
\\
b_{2}^{2}=E^{2}\Omega^{2}+e\Omega B_{0}E+M^{2}\omega^{2}\xi_{1}^{2}+\frac{e^{2}B_{0}^{2}}{4},\\
\\
b_{3}=M\lambda,\\
\\
b_{4}=E^{2}+\frac{\Omega e \Phi_{B}}{\alpha \pi}E-\frac{2\Omega l}{\alpha}E-2M\omega \xi_{1}-2M^{2}\omega^{2} \xi_{1}\xi_{2}-\frac{e B_{0} l}{\alpha}+\frac{e^{2}\Phi_{B} B_{0}}{2\pi \alpha}-k^{2}-M^{2}.
\end{array}\end{equation}
We define
\begin{equation}
s(r)=\exp(-\frac{1}{2}b_{2}r^{2})r^{\sqrt{b_{1}}}H(r),
\end{equation}
$H(r)$ is to be determined. Substituting the new transformation into the $Eq.(26)$, one gets
\begin{equation}
H^{''}+((2\sqrt{b_{1}}+1)\frac{1}{r}-2b_{2}r)H^{'}+(b_{4}-2b_{2}-2\sqrt{b_{1}}b_{2}-2b_{3}\frac{1}{r})H=0.
\end{equation}
Introducing the new variable $\rho=\sqrt{b_{2}}r$ into the above equation, the second order different equation about $\rho$ reads
\begin{equation}\begin{array}{l}
H^{''}+(\frac{1+2\sqrt{b_{1}}}{\rho}-2\rho)H^{'} +[\frac{b_{4}}{b_{2}}-2-2\sqrt{b_{1}}-\frac{2b_{3}}{\sqrt{b_{2}}}\frac{1}{\rho}]H=0.
\end{array}\end{equation}
This second order differential equation is called as the biconfluent Heun equation and the related function named the biconfluent Heun equation. Take same methods like the  previous part, using the Frobenius method, $H(\rho)$ can be shown by the power series expansion $(j=0)$
\begin{equation}
H(\rho)=\Sigma_{n=0}C_{n}\rho^{n}.
\end{equation}
In addition, we can obtain relations:
\begin{equation}
C_{n+1}=0, \\\  \frac{b_{4}}{b_{2}}-2-2\sqrt{b_{1}}=2n,
\end{equation}
where n is a positive integer number. Combination the second condition and $Eq.(27)$, we obtain
\begin{equation}\begin{array}{l}
(2n+2+2\sqrt{M^{2}\omega^{2}\xi_{2}^{2}+\lambda^{2}+\frac{1}{\alpha^{2}}\tilde{l}_{ef}^{2}}) \sqrt{E^{2}\Omega^{2}+e\Omega B_{0}E+M^{2}\omega^{2}\xi_{1}^{2}+\frac{e^{2}B_{0}^{2}}{4}}=E^{2}-2M\omega \xi_{1}-\\2M^{2}\omega^{2} \xi_{1}\xi_{2}-\tilde{l}_{ef}(\frac{2\Omega E}{\alpha}+\frac{eB_{0}}{\alpha})-k^{2}-M^{2}.
\end{array}\end{equation}
\begin{figure}
\begin{minipage}[t]{0.45\linewidth}
\centerline{\includegraphics[width=7.5cm]{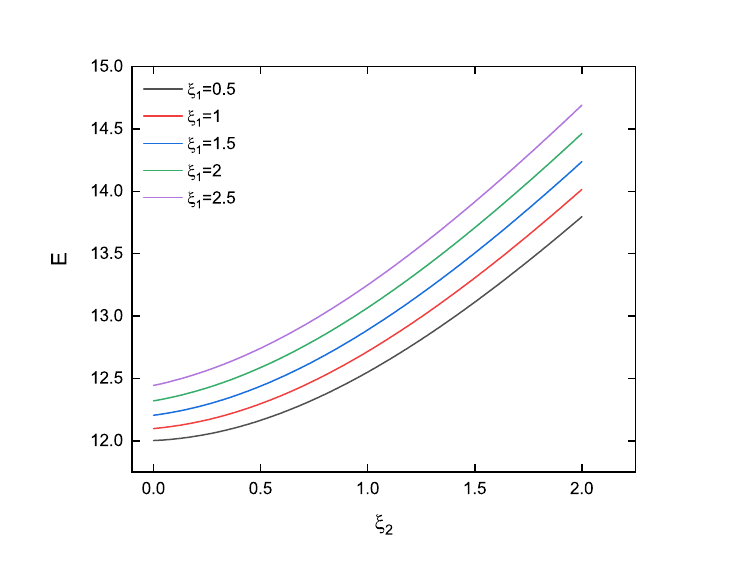}}
\centerline{(a)}
\end{minipage}
\begin{minipage}[t]{0.48\linewidth}
\centerline{\includegraphics[width=7.5cm]{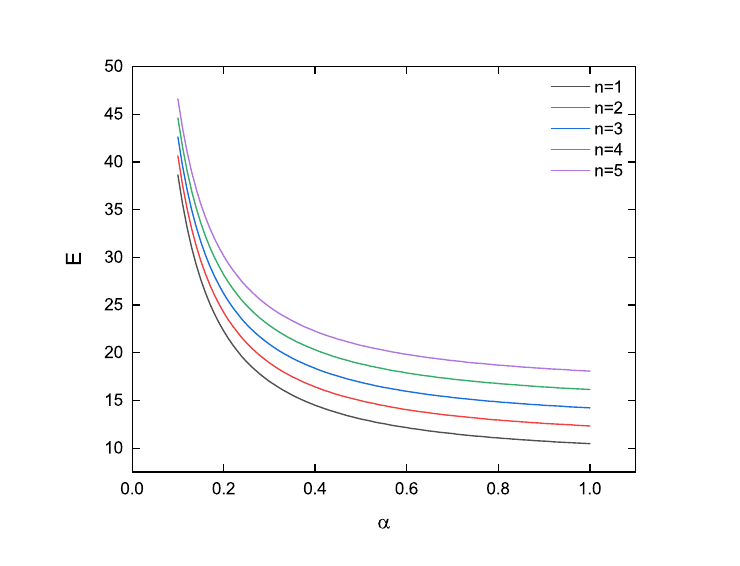}}
\centerline{(b)}
\end{minipage}
\parbox[c]{15.0cm}{\footnotesize{\bf Fig.~3.} (a) Energy eigenfunctions $E$ for different  Cornell potential parameter $\xi_{2}$ with values of the Cornell \\ parameter $\xi_{1}(0.5, 1, 1.5, 2, 2.5)$,$M=e=\Phi_{B}=n=\lambda=B_{0}=l=1$,$\omega=\Omega=k=1$, $\alpha=0.5$; \\ (b) Energy eigenfunctions $E$ for different angular
parameter $\alpha$ with values of the quantum \\ number n(1, 2, 3, 4, 5), $M=e=\Phi_{B}=\omega=\lambda=B_{0}=l=\Omega=k=\xi_{2}=1$, $\xi_{1}=2 $.}
\end{figure}
with the effective angular momentum $\check{l}_{ef}=l-\frac{e\Phi_{B}}{2\pi}$ is associated wit $\Phi_{B}$. Similarly,  the analogue of the Aharonov-Bohm effect can be given in \cite{t2,t3,D12,D13}. In addition, we consider to analyze the effect from the Cornell potential parameter  $\xi_{2}$ under the various values $\xi_{1}$ of to the positive energy eigenvalues by plots. We can observe that the energy is in risen trend with the Cornell potential parameter $\xi_{2}$ increase. And the higher the value of the $\xi_{1}$, the higher the corresponding curve. On the other plot, there is same trend with $Fig. 1$ $(a)$ but higher position of the curve than $Fig. 1$ $(a)$.

\section{Conclusion}
In this work we respectively investigate the K-G oscillator and the generalized K-G oscillator with Coulomb potential under the uniform magnetic in the Som-Raychaudhuri space-time and represent the implicit expressions of the energy under those cases, and analyze the analogue of the Aharonov-Bohm effect. In addition, we give the  effects of corresponding parameters on energy eigenvalues, by giving the some plots. Fig. 1 and Fig. 2 are about the energy eigenvalues of K-G oscillator with the considered parameters including the angular parameter $\alpha$ and oscillator frequency $\omega$, the vorticity parameter $\Omega$ and the Coulomb potential parameter $\lambda$. In $Fig. 1$ $(b)$, $Fig. 2$ $(a)$ and $Fig. 2$ $(b)$, there are all show increase trend of E, while E decrease with the rise values of the angular parameter $\alpha$ in $Fig. 1$ $(a)$. We separately consider the effects of the angular parameter $\alpha$ and Cornell potential parameter $\xi_{2}$ to the energy eigenvalues under the generalized K-G oscillator, in $Fig. 3$ $(b)$, the position of the curves are relatively higher than first case although both have same trend while there are the increase trend of energy eigenvalues with rise values of Cornell potential parameter $\xi_{2}$ under different $\xi_{1}$ in $Fig. 3$ (a) and the carve which the value of $\xi_{1}$ is bigger is higher. In addition, when  $\omega\rightarrow 0, \lambda\rightarrow 0, B_{0}\rightarrow 0$, the energy levels of the K-G oscillator is consistent with the result ($\xi\rightarrow 0,\xi_{c}\rightarrow 0 $) from the Ref\cite{t1} .

\section*{Acknowledgments}
This work is supported by the National Natural Science Foundation of China (Grant nos. 11465006 and 11565009) and the Major Research Project of innovative Group of Guizhou province (2018-013).\\





\end{document}